\documentclass[12pt]{article}
\usepackage{amsmath,amssymb}
\usepackage{graphicx}
\usepackage{wrapfig}

\usepackage{hyperref}
\usepackage{cite}

\def\beq{\begin{eqnarray}}
\def\eeq{\end{eqnarray}}

\newcommand{\Tr}{\,\mathrm{Tr}\,}            

\newcommand{\be}{\begin{equation}}
\newcommand{\ee}{\end{equation}}
\newcommand{\bea}{\begin{eqnarray}}
\newcommand{\eea}{\end{eqnarray}}
\newcommand{\bg}{\begin{gather}}
\newcommand{\eg}{\end{gather}}
\newcommand{\bseq}{\begin{subequations}}
\newcommand{\eseq}{\end{subequations}}

\def\be{\begin{eqnarray}}
\def\ee{\end{eqnarray}}
\def\lb{\label}

\begin{document}

\title{\textbf{Positive cosmological constant, non-local gravity  
and horizon entropy  }}
\vspace{2.5cm}
\author{ \textbf{
 Sergey N. Solodukhin$^\sharp$ }} 

\date{April 05, 2012}
\maketitle

\begin{center}
  \hspace{-0mm}
  \emph{ Laboratoire de Math\'ematiques et Physique Th\'eorique, }\\
  \emph{Universit\'e Fran\c cois-Rabelais Tours, F\'ed\'eration Denis Poisson - CNRS, }\\
  \emph{Parc de Grandmont, 37200 Tours, France} \\
\end{center}

{\vspace{-11cm}
\begin{flushright}
\end{flushright}
\vspace{11cm}
}



\begin{abstract}
\noindent { We discuss a class of (local and non-local) theories of gravity  that share same properties: i) they admit  the Einstein spacetime with arbitrary cosmological constant as a
solution; ii) the on-shell action of such a theory vanishes and iii) any (cosmological or black hole) horizon in the Einstein spacetime with a positive cosmological constant does not have a non-trivial entropy.   The main focus is made on a recently proposed non-local model. This model has two phases: with a positive cosmological constant $\Lambda>0$ and with zero $\Lambda$. The effective gravitational coupling differs essentially in these two phases. Generalizing the previous result of Barvinsky  we show that the non-local theory in question is free of ghosts on the background of any Einstein spacetime and that it propagates a standard spin-2 particle. Contrary to the phase with a positive $\Lambda$, where the  entropy vanishes for any type of horizon,  in an Einstein  spacetime with zero cosmological constant the horizons have the ordinary entropy proportional to the area.  
We conclude that, somewhat surprisingly,  the presence of any, even extremely tiny,  positive cosmological constant should be important for the proper resolution of the  entropy problem and, possibly, the 
information puzzle. }
\end{abstract}

\vskip 2 cm
\noindent
\rule{7.7 cm}{.5 pt}\\
\noindent 
\noindent
\noindent ~~~$^{\sharp}$ {\footnotesize e-mail: Sergey.Solodukhin@lmpt.univ-tours.fr}


\newpage

\section{ Introduction}
\setcounter{equation}0
When one first learns about the fact that the (black hole or cosmological)  horizons have certain entropy \cite{BH} proportional to the area the first natural question that 
is tempted to ask is whether this entropy is due to the well-known degrees of freedom already present in the theory, namely the degrees of freedom of a spin-2 particle
(a graviton)? A now popular (but not unique)  point of view is that the answer to this question is negative and there should be some other, new, degrees of freedom additional to that of a graviton that are there only if there is a horizon in the spacetime and that are otherwise  invisible.  In the approach of \cite{CFT}, which is the author's favorite, these degrees of freedom appear in the near-horizon region and are effectively described by a 2d conformal field theory.  However, even before one comes up with any idea of what the explanation for the entropy  should be it is rather clear
 that the degrees of freedom of a spin-2 particle should be excluded\footnote{For an alternative point of view see \cite{Dvali:2011aa}.} as such a universal explanation by a counter-example of gravity in three dimensions. Indeed, in three space-time dimensions  General Relativity, with or without cosmological constant, does not describe any propagating gravitational degrees of freedom. On the other hand, the BTZ black hole \cite{Banados:1992wn}, which is a solution in the theory with a negative cosmological constant, nevertheless has a non-vanishing entropy. 
This entropy can not be explained as (only) the entropy of  the propagating massless spin-2 particles since the latter do not exist in three dimensions.
 In this paper we consider a class of theories that may be viewed as a counter-example in the opposite direction. These theories have the certain  propagating gravitational degrees of freedom while the horizons do not have any non-trivial entropy. More precisely, the entropy of a horizon of any  type is zero. Thus, quantum mechanically, the relevant quantum state, responsible for the entropy,  is   a pure state.

The theories we are going to consider share the following same properties:

\noindent i) the Einstein spacetime 

\be
R_{\mu\nu}=\Lambda g_{\mu\nu}
\lb{1}
\ee
is a solution for arbitrary cosmological constant $\Lambda$. The fact that value of $\Lambda$ is not determined is a manifestation of the (direct or hidden) scale invariance
present in the theory.

\noindent ii) the on-shell gravitational action is identically zero for any value of $\Lambda$.

\noindent iii) if the Einstein spacetime (\ref{1}) contains a (non-extremal) horizon its entropy is zero.

We first analyze the local theories of gravity with the discussed  properties. These theories are quadratic in curvature and  suffer from the obvious sickness: they 
contain ghosts. Our second class of examples is motivated by the non-local model recently proposed by Barvinsky \cite{Barvinsky:2011hd}. This model has been shown in  \cite{Barvinsky:2011hd} to describe a maximally symmetric spacetime with arbitrary cosmological constant. Moreover, the theory is  ghost free on a maximally symmetric background. In this paper we demonstrate that there exists a more general solution  of the type (\ref{1}) that describes an Einstein spacetime  with arbitrary cosmological constant $\Lambda$. Then, we generalize the statement on the absence of the ghosts for this more general solution and, most importantly, we show that the horizon entropy in this class of theories 
vanishes for any type of horizon. Throughout the paper we work with the Euclidean theory  of gravity. The Lorentzian part of the story, although very important,
is not considered here.

\section{ Local theories quadratic in curvature}
\setcounter{equation}0
Let us start with a generic gravitational action quadratic in derivatives. In four dimensions the square of the Riemann tensor can be eliminated by using the Gauss-Bonnet theorem
and, thus, this action is a linear combination of two terms
\be
W=\int \sqrt{g}(a\, R^2+b\, R_{\mu\nu}R^{\mu\nu})\, ,
\lb{2}
\ee
where $a$ and $b$ are some coupling constants. First of all, we notice that this action is scale invariant, i.e. it does not change under the constant rescaling of the metric  
$g_{\mu\nu}\rightarrow \lambda g_{\mu\nu}$. As a consequence of this invariance, the gravitational equations obtained by varying (\ref{2})  with respect to metric admit
spacetime (\ref{1}) with arbitrary $\Lambda$ as a solution.  Indeed, the gravitational equations can be written in the form
\be
a\, E_{\mu\nu}R+b\, (E_{\mu}^{\ \alpha}E_{\alpha\nu}+\frac{1}{2} R E_{\mu\nu}-\frac{1}{4} g_{\mu\nu} E_{\alpha\beta}E^{\alpha\beta})+\nabla\nabla( {\cal R})=0\, ,
\lb{3}
\ee
where $E_{\mu\nu}=R_{\mu\nu}-\frac{1}{4}g_{\mu\nu}R$ is the traceless part of the Ricci tensor. The last term in (\ref{3}) contains the covariant derivatives of
the Ricci tensor and the Ricci scalar. For the Einstein spacetimes (\ref{1}) this term vanishes identically. The remaining terms in (\ref{3}) contain linear and quadratic combinations of tensor $E_{\mu\nu}$. This tensor vanishes, 
\be
E_{\mu\nu}=0\, ,
\lb{*}
\ee
 for the Einstein metrics (\ref{1}) with arbitrary $\Lambda$.  The Einstein metrics thus are solutions in the generic quadratic theory
(\ref{2}). 

Let us now restrict the class of theories by imposing the condition  that the on-shell action, i.e. the action (\ref{2}) considered on the metrics (\ref{1}),
should vanish. This condition puts constraint  on the couplings $a$ and $b$, namely one has that $a=-b/4$. The new class of theories is described by the action
\be
W_{N}=b\int(R_{\mu\nu}R^{\mu\nu}-\frac{1}{4}R^2)=b\int E_{\mu\nu}E^{\mu\nu}\, ,
\lb{4}
\ee
quadratic in tensor $E_{\mu\nu}$.

Now, let us turn to the entropy. Among the  Einstein spacetimes (\ref{1}) there are  those with horizons. If the cosmological constant $\Lambda$ is positive then the obvious example is the Schwarzschild-de Sitter spacetime which has horizons of two types, cosmological and black hole. So what is the entropy of any of these horizons in the theory (\ref{4})?
Off-shell, for an arbitrary metric with a horizon 2-surface $\Sigma$, the horizon entropy in the theory (\ref{4}) equals (see \cite{Fursaev:1995ef} for the detail explanations)
\be
S=-4\pi b \int_\Sigma E_{ii}\, ,
\lb{5}
\ee
where $E_{ii}=E_{\mu\nu}n^\mu_i n^\nu_i$ and  $n^\mu_i$, $i=1,2$ are two vectors normal to $\Sigma$.  Then, on-shell, for any Einstein metric (\ref{1}), the entropy (\ref{5}) vanishes, $S=0$,
due to (\ref{*}).

Assuming that the first law is still valid one would expect that the energy in the theory (\ref{4}) should vanish. This is indeed the case as was shown earlier\footnote{The author 
thanks B. Tekin for bringing this reference to him.}  in \cite{Deser:2002jk}.

Any theory of the type (\ref{2}) is sick due to the presence of a ghost in the spectrum. Usually, the ghost appears in the propagator as a $1/(p^2+m_1^2)$ pole with a negative residue.
Another pole $1/(p^2+m_2^2)$, with a positive residue, corresponds to a physical particle.  The value of the residue at a pole gives the value of the gravitational coupling.
 The theory (\ref{4}) is special since in this case the two poles collide $(m_1=m_2=m$) and form a single $1/(p^2+m^2)^2$ pole.
In a covariant way this can be seen by looking at a small perturbations $h$ on the background of the  Einstein metric (\ref{1}).  The gravitational action (\ref{2}) for the transverse-traceless perturbation  $h$ then takes the form \cite{GV}
\be
W=\frac{b}{2}\int h (\Delta_L+\frac{1}{2}R)(\Delta_L+(1+2t)R)h\, ,
\lb{6}
\ee
where $(\Delta_L h)_{\mu\nu}=\Delta h_{\mu\nu}+2R_{\mu\alpha\nu\beta}h^{\alpha\beta}-\frac{1}{4} Rh_{\mu\nu}$ is the Lichnerowicz operator ($\Delta=\nabla_\alpha\nabla^\alpha$)
and we defined $t=-a/b$. The theory (\ref{4}) corresponds to  $t=-1/4$, this is the only case when the quartic operator in (\ref{6}) is the square of a second order operator,
\be
W_N(h)=\frac{b}{2}\int h(\Delta_L+\frac{1}{4}R)^2h\, .
\lb{7}
\ee

The theories quadratic in curvature have been recently studied in \cite{tHooft:2011aa} and \cite{Maldacena:2011mk}, the focus however was mostly made on the conformally invariant  theory 
which corresponds to the action (\ref{2}) with $t=-1/3$.

\section{ Non-local theories of gravity}
\setcounter{equation}0
The non-local theories of gravity which we are going to discuss in this section are inspired by a recent interesting work of Barvinsky \cite{Barvinsky:2011hd}.
Let us, however, start with a somewhat simpler model which involves only the Ricci scalar.

\subsection{A simple non-local model expressed  in terms of Ricci scalar only}
\lb{sec:scalar}

In the previous section we did not include in the action the Einstein-Hilbert term proportional to the Ricci scalar $R$. This term would clearly break the scale invariance 
and spoil the properties of the theory discussed in the section 2. Nevertheless, it is possible to have a theory with  the Einstein-Hilbert term that would still possess the properties i)-iii)   
provided some non-local terms are added to the action. Let us consider a theory described by the following gravitational action (note that we use the units $16\pi G=1$)
\be
W=\int \sqrt{g}\left(-R-\alpha R \,\frac{1}{\Delta-\alpha R}\, R\right)\, ,
\lb{8}
\ee where $\Delta=\nabla_\mu \nabla^\mu$ is the scalar Laplace operator, and $\alpha$ is a parameter. The earlier work on the non-local modifications of gravity includes \cite{ArkaniHamed:2002fu}, \cite{Barvinsky:2003kg},  \cite{Deser:2007jk}, \cite{Nojiri:2007uq}. 

At first sight, the theory (\ref{8}) appears to have  two different phases, depending on what term in the  differential operator $(\Delta -\alpha R)$ is dominating. If the spacetime curvature  is small compared to the second derivatives, $R\ll \nabla\nabla$, then $R$ can be neglected in the operator  and the second, non-local, term in the action (\ref{8}) is a small correction to the Einstein-Hilbert term, provided $\alpha$ is sufficiently small. On the other hand, if the curvature is large compared to the second derivatives, $R\gg \nabla\nabla$, then we can neglect the Laplace operator so that  the second term in (\ref{8}) becomes local and exactly cancels the Einstein-Hilbert term. In fact this qualitative analysis is somewhat misleading. We will see that the presence of any, even extremely tiny, positive cosmological constant affects  the dynamics of the theory  on all scales. So the phases just discussed are realized in different spacetimes
with the exactly vanishing and non-vanishing Ricci scalar.

The non-locality in (\ref{8}) can be localized by introducing an extra scalar field $\phi$ with the action
\be
W_\phi=-\int\sqrt{g} \left(\alpha(\nabla\phi)^2+2\alpha R\phi+\alpha^2R\phi^2\right)\, .
\lb{phi}
\ee
Variation with respect to $\phi$ gives the field equation
\be
(\Delta-\alpha R)\phi=R\, .
\lb{9}
\ee
Solving this equation for $\phi$ and substituting back to (\ref{phi}) gives exactly the non-local term in (\ref{8}).
So that the total action (\ref{8}) takes the local form
\be
W=\int \sqrt{g}\left( -R(1+\alpha\phi)^2-\alpha(\nabla\phi)^2\right)\, .
\lb{8-1}
\ee
The gravitational equations, which one obtains by varying the action (\ref{phi}) with respect to metric, then take the form
\be
&&-(1+\alpha\phi)^2(R_{\mu\nu}-\frac{1}{2}g_{\mu\nu}R)-\alpha\partial_\mu\phi\partial_\nu\phi +\frac{\alpha}{2}g_{\mu\nu}(\nabla\phi)^2\nonumber \\
&&-\nabla_\mu\nabla_\nu (1+\alpha\phi)^2+g_{\mu\nu}\Delta (1+\alpha\phi)^2=0\, .
\lb{10}
\ee
By taking the trace of this equation we arrive at 
\be
R(1+\alpha \phi)^2+\alpha(\nabla\phi)^2+ 3\Delta(1+\alpha\phi)^2=0\, .
\lb{11}
\ee

\medskip

\noindent Consider two different cases.

\medskip

\noindent {\it A. Ricci scalar $R=0$.}
In this case the equation (\ref{9}) for the auxiliary field $\phi$ and the gravitational constraint (\ref{11})  take, respectively, the form 
\be
\Delta\phi=0\, , \  \ \ \alpha(1+6\alpha)(\nabla\phi)^2=0\, .
\lb{12}
\ee
The only possible solution to these two equations is constant, 
\be
\phi=const\, .
\lb{14}
\ee
Substituting this  to gravitational equations (\ref{10}), we find that 
\be
R_{\mu\nu}=0\, ,
\lb{15}
\ee
i.e. the spacetime is Ricci flat. 

In particular, the Schwarzschild metric is a possible solution. It has a horizon surface $\Sigma$.
The gravitational entropy of the horizon, 
\be
S=4\pi\int_\Sigma (1+\alpha\phi)^2\, ,
\lb{16}
\ee
is non-zero for a generic value of constant $\phi$. In this phase the theory (\ref{8}), (\ref{8-1}) is more like the usual Einstein-Hilbert theory of gravity.

\medskip

\noindent {\it B. Ricci scalar $R\neq 0$.} 
In this case we can decompose the auxiliary field $\phi$  as follows
\be
\phi=-\frac{1}{\alpha}+\phi_0\, ,
\lb{17}
\ee
where $\phi_0$ satisfies the homogeneous equation
\be
(\Delta-\alpha R)\phi_0=0\, .
\lb{18}
\ee
The constraint (\ref{11}) then  takes the form
\be
(1+6\alpha)\phi_0^2R+(\alpha+6)(\nabla\phi_0)^2=0\, .
\lb{19}
\ee
We are, in particular, interested in the case, when $\alpha$ is a small parameter so that both $(1+6\alpha)$ and $(\alpha+6)$ are positive. Suppose now that the Ricci scalar is a positive definite function, $R(x)>0$. Then, the equation (\ref{19}) has only the trivial solution
\be
\phi_0(x)=0\, .
\lb{20}
\ee  
We may come to the same  conclusion by looking just at the field equation (\ref{18}).  Indeed, let us multiply (\ref{18}) by $\phi_0$ and integrate. Provided the Euclidean manifold $M$ is
closed, so that there is no a boundary term after integration by parts, we get that
\be 
\int_M\left((\nabla\phi_0)^2+\alpha R\phi_0^2\right)=0\, .
\lb{21}
\ee
For positive $\alpha$ and provided the Ricci scalar of manifold $M$ is everywhere positive, $R(x)>0$, we find that there exists only the trivial solution   (\ref{20}).
The gravitational equations (\ref{10}) then are automatically satisfied not imposing any new constraint on the metric of manifold $M$. We note that if there are regions, where the Ricci scalar is negative, $R(x)<0$, there can be non-trivial solutions both of the field equation (\ref{18}) and the constraint (\ref{19}).
It is only in the regions, where the Ricci scalar is positive, the metric is not fixed by the field equations and the auxiliary field $\phi_0(x)$ is frozen at the trivial value (\ref{20}). If there is a horizon in one of the regions, where $R(x)>0$, then its entropy, given by (\ref{16}), vanishes, $S=0$.

\subsection{Non-local model of Barvinsky}

{\it The model.} In the model recently  proposed in \cite{Barvinsky:2011hd} one adds a non-local term to the standard Einstein-Hilbert gravitational action\footnote{Note that our definition of $\alpha$ in the action (\ref{22}) differs by factor of 4 from that of used in \cite{Barvinsky:2011hd}.}
\be
&&W=-\int \sqrt{g} R +W_B\, , \nonumber \\
&&W_B=4\alpha\int \sqrt{g}R^{\mu\nu}\frac{1}{\Delta+\hat{P}}G_{\mu\nu}\, ,
\lb{22}
\ee
where $G_{\mu\nu}=R_{\mu\nu}-\frac{1}{2}g_{\mu\nu}R$ and the symmetric matrix $\hat{P}$ contains terms linear in curvature
\be
\hat{P}\equiv P_{\alpha\beta}^{\ \ \mu\nu}=aR_{(\alpha\ \beta )}^{\ (\mu\ \nu )}+b(g_{\alpha\beta}R^{\mu\nu}+g^{\mu\nu}R_{\alpha\beta})+cR_{(\alpha}^{(\mu}\delta_{\beta )}^{\nu )}+dRg_{\alpha\beta}g^{\mu\nu}+eR\delta^{\mu\nu}_{\alpha\beta}\, .
\lb{23}
\ee
This matrix has a nice property that
\be
\hat{P}g_{\mu\nu}\equiv P_{\mu\nu}^{\ \ \alpha\beta}g_{\alpha\beta}=AR_{\mu\nu}+Bg_{\mu\nu}R\, ,
\lb{24}
\ee
where
\be
A=a+4b+c\, , \,
&&B=b+4d+e\, .
\lb{25}
\ee
In the model of Barvinsky one fine tunes the values of parameters $A$ and $B$ in the matrix $\hat{P}$ (\ref{24}) and the value of the coupling $\alpha$,
\be
B+\frac{A}{4}=-\alpha\, .
\lb{30}
\ee
Let us first slightly rewrite the action by separating the trace and traceless parts in the Ricci tensor. Then, using that
\be
&&R_{\mu\nu}=E_{\mu\nu}+\frac{1}{4}g_{\mu\nu}R\, ,\nonumber  \\
&&G_{\mu\nu}=E_{\mu\nu}-\frac{1}{4}g_{\mu\nu}R\, 
\lb{26}
\ee
the Barvinsky's term in (\ref{22}) can be rewritten as
\be
W_B=4\alpha \int\sqrt{g}\left(-\frac{1}{16}Rg^{\mu\nu}\frac{1}{\Delta+\hat{P}}\, g_{\mu\nu}R+E^{\mu\nu}\frac{1}{\Delta+\hat{P}}E_{\mu\nu}\right)\, .
\lb{27}
\ee

\medskip

\noindent {\it The local form of the action.} 
This non-local action can be rewritten in a local form by introducing the auxiliary tensor fields $\phi_{\mu\nu}$ and $\psi_{\mu\nu}$ as follows
\be
W_B=4\alpha\int \left(\phi^{\mu\nu}(\Delta+\hat{P})\phi_{\mu\nu}-\frac{1}{2}\phi R-\psi^{\mu\nu}(\Delta+\hat{P})\psi_{\mu\nu}+2\psi^{\mu\nu} E_{\mu\nu}\right)\, .
\lb{B-local}
\ee
The equation of motion for the auxiliary field $\phi_{\mu\nu}$ is
\be
(\Delta +\hat{P})^{\ \ \alpha\beta}_{\mu\nu}\phi_{\alpha\beta}=\frac{1}{4}g_{\mu\nu}R\, .
\lb{28}
\ee
It is useful to make a decomposition on the trace and the traceless parts, $\phi_{\mu\nu}=\frac{1}{4}g_{\mu\nu}\phi+\tilde{\phi}_{\mu\nu}$, $g^{\mu\nu}\tilde{\phi}_{\mu\nu}=0$.
Respectively, separating the trace and traceless parts of the equation (\ref{28}) and using (\ref{30}) we obtain the equations
\be
(\Delta\phi -\alpha R\phi)=R-AE^{\alpha\beta}\tilde{\phi}_{\alpha\beta}\, , \ \ (\Delta +\hat{P})^{\ \ \alpha\beta}_{\mu\nu}\tilde{\phi}_{\alpha\beta}=-\frac{1}{4}A E_{\mu\nu}\phi+\frac{1}{4}Ag_{\mu\nu}E^{\alpha\beta}\tilde{\phi}_{\alpha\beta}\, .
\lb{29}
\ee
We note that 
only if the traceless part of the Ricci tensor vanishes, $E_{\mu\nu}=0$, the two equations in (\ref{29}) disentangle,
\be
(\Delta-\alpha R)\phi=R\, , \ \ \   (\Delta +\hat{P})^{\ \ \alpha\beta}_{\mu\nu}\tilde{\phi}_{\alpha\beta}=0\, .
\lb{29-1}
\ee
This is exactly the case of the Einstein spacetime (\ref{1}).
  
Similarly, the equation of motion for the auxiliary field $\psi_{\mu\nu}$ is
\be
(\Delta +\hat{P})^{\ \ \alpha\beta}_{\mu\nu}{\psi}_{\alpha\beta}= E_{\mu\nu}\, .
\lb{32}
\ee
Representing $\psi_{\mu\nu}$ as a sum of the trace and traceless parts, $\psi_{\mu\nu}=\frac{1}{4}g_{\mu\nu}\psi+\tilde{\psi}_{\mu\nu}$, $g^{\alpha\beta}\tilde{\psi}_{\alpha\beta}=0$, and separating the trace and traceless parts in the equation (\ref{32}) we find that
\be
(\Delta-\alpha R)\psi=-AE^{\alpha\beta}\tilde{\psi}_{\alpha\beta}\, , \ \  (\Delta +\hat{P})^{\ \ \alpha\beta}_{\mu\nu}\tilde{\psi}_{\alpha\beta}=E_{\mu\nu}+\frac{1}{4}A
g_{\mu\nu} E^{\alpha\beta}\tilde{\psi}_{\alpha\beta}-\frac{1}{4}A E_{\mu\nu}\psi\, .
\lb{33}
\ee
Only for  the Einstein spacetime, $E_{\mu\nu}=0$, the trace and traceless parts decouple in equations (\ref{33}) and we have that
\be
 (\Delta-\alpha R)\psi=0\, , \ \ \ (\Delta +\hat{P})^{\ \ \alpha\beta}_{\mu\nu}\tilde{\psi}_{\alpha\beta}=0\, .
\lb{34}
\ee

The total gravitational action (\ref{22}) can be written in the local form in terms of the trace and traceless parts of the auxiliary fields
\be
&&W=\int [-(1+\alpha\phi)^2R+\alpha\phi\Delta\phi +4\alpha\tilde{\phi}^{\mu\nu}(\Delta+\hat{P})\tilde{\phi}_{\mu\nu}+2\alpha A\tilde{\phi}^{\mu\nu}E_{\mu\nu}\phi\nonumber \\
&& -\alpha\psi\Delta\psi +\alpha^2 R\psi^2 -4\alpha \tilde{\psi}^{\mu\nu}(\Delta+\hat{P})\tilde{\psi}_{\mu\nu} +8\alpha\tilde{\psi}^{\mu\nu}E_{\mu\nu}-2\alpha A\psi E_{\mu\nu}\tilde{\psi}^{\mu\nu}]\, .
\lb{W-local}
\ee
The part of the action (\ref{W-local}) that depends only on the field $\phi$ is exactly the action (\ref{8-1}) considered in the section 3.1.

Yet another useful form of the gravitational action is obtained by substituting the solutions of the equations (\ref{28}) and (\ref{32}) back to the action,
\be
W= \int \sqrt{g}  \left(-(1+\alpha \phi)R+4\alpha E^{\mu\nu}\tilde{\psi}_{\mu\nu}\right)\, .
\lb{local}
\ee
Note, that in this form of the action  the auxiliary fields $\phi$ and $\tilde{\psi}_{\mu\nu}$ are not independent from the metric  as they are solutions to equations (\ref{29}) and (\ref{33}).
We notice that even though the components $\tilde{\phi}_{\mu\nu}$ and $\psi$ do not appear in the action (\ref{local}) they do show up in the gravitational equations, when the variation of (\ref{local}) with respect to metric is computed,  via the equations (\ref{29}) and (\ref{33}).

\medskip

\noindent {\it Einstein spacetime ($E_{\mu\nu}=0$) as a solution and the uniqueness.}
For an Einstein metric the variation of the second term in (\ref{27}) (or in (\ref{local})) with respect to metric is not a priori vanishing. It contains the covariant derivatives acting on the traceless tensor $\tilde{\psi}_{\alpha\beta}$ which satisfies the  second equation in (\ref{34}). There always exists the trivial solution to this equation, $\tilde{\psi}_{\alpha\beta}=0$,
for which the respective contribution to the gravitational equations vanishes. The theory then effectively reduces to the theory with only the Ricci  scalar non-locality that was analyzed in section 3.1. The remaining gravitational equations are expressed in terms of the scalar $\phi$  which satisfies equation (\ref{29-1}) and  are automatically satisfied provided $\phi=-\frac{1}{\alpha}$. Thus, there always (for any choice of couplings in operator $\hat{P}$ provided the constraint (\ref{30}) is satisfied)  exists the
solution describing an Einstein spacetime (\ref{1}) with arbitrary cosmological constant $\Lambda$,
\be
E_{\mu\nu}=0\, , \ \ \tilde{\psi}_{\mu\nu}=0\, , \ \ \phi=-\frac{1}{\alpha}\, , \ \ \psi=0\, , \ \ \tilde{\phi}_{\mu\nu}=0\, .
\lb{35}
\ee
The gravitational action (\ref{local}) vanishes for this solution, $W=0$.
This result is more general than the one presented in \cite{Barvinsky:2011hd}, where only the maximally symmetric solution was discussed. Here
we showed that the solutions describing an Einstein spacetime  (\ref{1}) exist for any choice of the couplings (subject to condition (\ref{30})) in operator  (\ref{23}).
The important question now is whether the solution (\ref{35}) is unique. By uniqueness we mean the existence of only the trivial solution for  the auxiliary fields (\ref{35}) provided the metric is Einstein. For the scalar $\phi$ this question was already discussed in section 3.1, where we concluded that if the Euclidean spacetime is a compact manifold with a sign definite Ricci scalar, $R>0$, and provided the coupling $\alpha>0$, there exists   only a constant solution  $\phi=-1/\alpha$ of the equation (\ref{29-1}) for $\phi$.
By same arguments there exists only the trivial solution $\psi=0$ for the trace part of $\psi_{\mu\nu}$.

Thus, it remains to analyze the uniqueness problem for the traceless tensor field $\tilde{\psi}_{\mu\nu}$ (and $\tilde{\phi}_{\mu\nu}$) satisfying equation (\ref{34}). For an Einstein metric (\ref{1}) the matrix
$\hat{P}$ (\ref{23}) reduces to the form
\be
\hat{P}_{\alpha\beta}^{\ \ \mu\nu}=aW_{(\alpha\ \beta)}^{\ (\mu \ \nu )}-\alpha \Lambda g_{\alpha\beta}g^{\mu\nu}-C\Lambda (\delta_{\alpha\beta}^{\mu\nu}-\frac{1}{4}g_{\alpha\beta}g^{\mu\nu})\, ,
\lb{36}
\ee
where $W_{\alpha\ \beta}^{\ \mu \ \nu } $ is the Weyl tensor and we introduced 
\be
C=-\frac{a}{3}-c-4e
\lb{37}
\ee
as in \cite{Barvinsky:2011hd}.
 Multiplying the second equation  in (\ref{34})
by $\tilde{\psi}^{\alpha\beta}$ and integrating by parts (the Euclidean manifold is assumed to have no boundaries), we find that
\be\int_M \left(\nabla^\mu\tilde{\psi}^{\alpha\beta}\nabla_\mu\tilde{\psi}_{\alpha\beta}-a\tilde{\psi}^{\alpha\beta}\, W_{\alpha \ \beta}^{\ \mu\ \nu}\, \tilde{\psi}_{\mu\nu}+C\Lambda \tilde{\psi}^{\alpha\beta}\tilde{\psi}_{\alpha\beta}\right)=0\, ,
\lb{38}
\ee
where we took into account that the tensor $\tilde{\psi}_{\alpha\beta}$ is traceless. 

Let us first consider the case when the 4-manifold is maximally symmetric de Sitter space or, in the Euclidean signature, the round sphere $S^4$. The Weyl tensor vanishes in this case.
The equation (\ref{38}) then implies that if constant $C>0$ and the cosmological constant is positive,
$\Lambda>0$, then there exists only the trivial solution to the field equation $(\Delta +\hat{P})_{\mu\nu}^{\ \ \alpha\beta}\tilde{\psi}_{\alpha\beta}=0$,
\be
\tilde{\psi}_{\alpha\beta}=0\, .
\lb{39}
\ee
Same arguments of course apply for the field $\tilde{\phi}_{\alpha\beta}$.   Thus, for the round sphere $S^4$  the solution (\ref{35}) is unique.

What can we say in the case when the Weyl tensor is non-vanishing, for instance if the Einstein 4-manifold is a deformation of the round sphere?
By the same arguments as above one can show that an eigen value $\lambda$ of the operator $-(\Delta +\hat{P})$ on the round sphere $S^4$ satisfies the inequality $\lambda\geq  C\Lambda=CR/4$. Thus, the minimal eigen value $\lambda_0\geq C\Lambda=CR/4$ is a strictly positive number. Suppose that the 4-manifold in question is a small deformation of the round sphere so that in an appropriate norm $\left\| W \right\|$ the Weyl tensor is small. Then the minimal eigen value of the operator $-(\Delta +\hat{P})$ is equal to $\lambda_0'=\lambda_0+\delta \lambda_0$, where $\delta \lambda_0 =-a\int_{M^4} \tilde{\psi}_0W\tilde{\psi}_0$ is small. So that $\lambda'_0$ is still strictly positive. We conclude that if the Weyl tensor is  small 
the zero mode of operator $-(\Delta +\hat{P})$  is trivial (\ref{39}). Thus, for small deformations of the round sphere we still have the uniqueness of the solution (\ref{35}).
Whether the uniqueness holds for a large perturbation or, actually, for any perturbation of the round sphere requires a more involved analysis. Possibly, this would require 
certain restrictions on the value of the parameter $a$ in the matrix $\hat{P}$. We leave this analysis outside the present paper.

In the case of negative cosmological constant ($\Lambda<0$), provided the constant $C>0$, our arguments based on equation (\ref{38}) can not be used to conclude the uniqueness of the
solution. This is so even for a maximally symmetric hyperbolic spacetime. The two terms in the  eq.(\ref{38}) then are not of the same sign and thus there could be a non-trivial solution for $\tilde{\psi}_{\alpha\beta}$. In section 3.1 we arrived at a similar conclusion for the scalar field $\phi$. Thus, most likely, for a negative cosmological constant the uniqueness,    we discuss here, does not take place. The positive sign of the cosmological constant is, hence, indispensable for the property of the uniqueness 
that plays the key role in the analysis presented in this paper.

\medskip

\noindent {\it Horizon entropy.}  In any theory of gravity described by  (local or non-local) action $W$ the horizon entropy  can be computed by the method of \cite{Fursaev:1995ef} as follows\footnote{In the approach of \cite{Fursaev:1995ef} the entropy is defined as response of the gravitational action to introduction of a conical singularity at the horizon. The metric itself then is a regular quantity while the curvature contains a delta-like singularity at the horizon  that contributes to the entropy.}
\be
S=-2\pi \int_\Sigma \frac{\delta W}{\delta R^{\mu\nu\alpha\beta}} \left((n^\mu n^\alpha)(n^\nu n^\beta)-(n^\mu n^\beta)(n^\nu n^\alpha)\right)\, ,
\lb{40}
\ee 
where $\Sigma$ is the horizon 2-surface, $n^\alpha_i$, $i=1,2$ are two vectors normal to $\Sigma$, $n_i^\mu n_j^\nu g_{\mu\nu}=\delta_{ij}$, and we defined
$(n^\alpha n^\beta)=\sum_{i=1}^2 n^\alpha_i n^\beta_i$. The entropy (\ref{40}) is an off-shell quantity valid for any metric not necessarily satisfying the gravitational equations.
On-shell, it is equivalent to  the Wald's entropy \cite{Wald:1993nt}. Thus, our strategy is to first obtain a general expression for the entropy (\ref{40}) in the theory (\ref{22}) and then consider this entropy for the solution (\ref{35}). 

The local form (\ref{W-local}) of the gravitational  action is the most convenient for the purposes of the entropy calculation. The auxiliary fields  in this action are considered to be
independent of the metric and, moreover, the action is linear in the curvature.  
We then obtain an off-shell expression for the horizon  entropy in the theory described by the action (\ref{W-local}),
\be
&&S=4\pi \int_\Sigma (1+\alpha\phi)^2 -4\pi\alpha^2\int_\Sigma\psi^2  -4\pi\alpha\int_\Sigma(A\phi\tilde{\phi}_{ii}-A\psi\tilde{\psi}_{ii}+4\tilde{\psi}_{ii})\lb{entropy} \\
&&-8\pi\alpha\int_\Sigma \left(a(\tilde{\phi}_{ii}^2-\tilde{\psi}_{ii}^2-\tilde{\phi}_{ij}^2+\tilde{\psi}^2_{ij})-c(\tilde{\phi}_{i\alpha}\tilde{\phi}_{i}^{\ \alpha}- \tilde{\psi}_{i\alpha}\tilde{\psi}_{i}^{\ \alpha})+2e(\tilde{\phi}^2_{\alpha\beta}-\tilde{\psi}^2_{\alpha\beta})\right)  \, ,\nonumber
\ee
where $\tilde{\phi}_{ij}=\tilde{\phi}_{\alpha\beta} n^\alpha_i n^\beta_j$, $\tilde{\phi}_{ii}=\sum_i\tilde{\phi}_{ii}$ and  $\tilde{\phi}_{i\alpha}=\tilde{\phi}_{\alpha\beta} n^\beta_i$.
On-shell,  for any solution (\ref{35}), which describes an Einstein spacetime\footnote{In the the Schwarzschild-de Sitter spacetime the two horizons,  black hole and cosmological, are present. It is well known that by choosing appropriately the periodicity in the Euclidean time one can make the manifold regular near one of the horizons but not, in general,  near both. There always appears a conical singularity at one of the horizons. The presence of the conical singularity does not change our conclusions on the uniqueness of the solution
since in the equation (\ref{38}) the possible boundary term at the singularity always vanishes provided the Friedrichs self-adjoint
extension \cite{Kay:1990cr} is chosen in which  the fields are regular at the conical singularity.} 
 with a positive cosmological constant $\Lambda$, the horizon entropy  (\ref{entropy}) 
  is  equal to zero,
\be
S=0\, .
\lb{41}
\ee
This is our main result. Thinking of possible generalizations, we note that for this result to be valid we would only need  the fields $\tilde{\psi}_{\alpha\beta}$, $\psi$, .. to vanish at the horizon surface $\Sigma$.
Thus, it is possible that one may relax our restrictions on the sign of the cosmological constant $\Lambda$ and on the values of the parameters in the action (\ref{22}).
This certainly opens various directions to further generalize our result. 

The non-locality of the original gravitational action (\ref{22}) is reflected in the fact that the solution to the gravitational equations, in general, is characterized by not just the metric but also by the auxiliary fields satisfying  the homogeneous equations.  
The uniqueness of the solution (\ref{35}) for positive $\Lambda$ guarantees that the auxiliary fields, classically,  do not produce any new degrees of freedom other than metric.  It is important for the universality of the property (\ref{41}), valid for any solution
that describes an Einstein spacetime. 
For a negative cosmological constant, even though we still have the solution (\ref{35}) for which the horizon entropy vanishes, this solution may not be unique.
Thus, there could be a more general solution which describes the Einstein spacetime (\ref{1})  with $\Lambda<0$ and with non-trivial auxiliary fields not necessarily vanishing at the horizon.  

\medskip

\noindent {\it No-ghosts in quadratic order.} In this part of our analysis we like to use the form (\ref{local}) of the gravitational action. Consider  a small perturbation $h_{\mu\nu}$ on the background of the solution (\ref{35}). We want to decompose the gravitational action
(\ref{local}) up to second order in perturbation.  
Due to triviality of the solution (\ref{35}) many terms in the decomposition  of (\ref{local}) in powers of the perturbation $h$ are vanishing and we have that
\be
\delta_2 W=\int\sqrt{g}\left(\frac{1}{2}\Tr h \, \delta_1\phi R-\alpha\delta_1\phi \delta_1 R  -\alpha R\delta_2\phi+4\alpha \delta_1E^{\mu\nu}\delta_1\tilde{\psi}_{\mu\nu}\right)\, ,
\lb{42}
\ee
where $\delta_1 $ contains terms linear in perturbation $h$ and $\delta_2$ contains terms quadratic in $h$, so that the total perturbation reads $\delta \phi=\delta_1\phi+\delta_2\phi+..$, where $\delta_n\phi\sim h^n$.  The term $\Tr h=g^{\mu\nu}h_{\mu\nu}$ comes from the variation of $\sqrt{g}$. Varying  the first equation in (\ref{29}) we find that $\delta_1\phi$ satisfies the homogeneous equation $(\Delta-\alpha R)\delta_1 \phi=0$ and, hence, vanishes, $\delta_1\phi=0$.  The first two terms in (\ref{42}), thus, vanish.
The quadratic variation $\delta_2\phi$ then satisfies equation
\be
(\Delta-\alpha R)\delta_2\phi=-A\delta_1 E^{\mu\nu}\delta_1\tilde{\phi}_{\mu\nu}\, ,
\lb{43}
\ee
where the variation $\delta_1\tilde{\phi}_{\mu\nu}$ is found from the  equation
\be
(\Delta+\hat{P})\delta_1\tilde{\phi}_{\mu\nu}=\frac{A}{4\alpha}\delta_1 E_{\mu\nu}\, .
\lb{44}
\ee
On the other hand, by  varying the second   equation in (\ref{33}) we obtain that $\delta_1\tilde{\psi}_{\mu\nu}$ satisfies 
\be
(\Delta+\hat{P})\delta_1\tilde{\psi}_{\mu\nu}=\delta_1E_{\mu\nu}\, .
\lb{45}
\ee
Inserting (\ref{43})-(\ref{45}) into (\ref{42}) we arrive at  the quadratic action
\be
\delta_2 W=(-\frac{A^2}{4\alpha}+4\alpha)\int \delta_1 E^{\mu\nu}\frac{1}{\Delta+\hat{P}}\delta_1 E_{\mu\nu}\, .
\lb{46}
\ee
The variation of the traceless tensor $E_{\mu\nu}$ results in
\be
\delta_1 E_{\mu\nu}=-\frac{1}{2}\hat{D} \bar{h}_{\mu\nu}\, , \ \ \hat{D}_{\mu\nu}^{\ \ \alpha\beta}=(\Delta -
\frac{R}{6})\delta_{\mu\nu}^{\alpha\beta}+2W^{(\alpha \ \beta )}_{\ (\mu \ \nu)}\, ,
\lb{47}
\ee
where $\bar{h}_{\mu\nu}=h_{\mu\nu}-\frac{1}{4}g_{\mu\nu}\Tr h$ is the traceless part of $h_{\mu\nu}$ and we imposed the gauge condition $\nabla_\alpha h^{\alpha}_\beta=\frac{1}{2}\partial_\beta \Tr h$. 
We note that operator $\hat{D}=\Delta_L+\frac{1}{4}R$, where $\Delta_L$ is the Lichnerowicz operator.  Now we introduce an effective Planck mass 
\be
\frac{M^2_{eff}}{2}=\frac{A^2}{4\alpha}-4\alpha=\frac{16B(B+2\alpha)}{\alpha}\, 
\lb{48}
\ee
in accordance with \cite{Barvinsky:2011hd}.
With this definition the quadratic part of the action reads
\be
\delta_2W=-\frac{M^2_{eff}}{2}\int \frac{1}{4}\bar{h}^{\mu\nu}\hat{D}\frac{1}{(\Delta+\hat{P})}\hat{D}\, \bar{h}_{\mu\nu}\, .
\lb{49}
\ee
Let us compare the structure of operators $\hat{D}$ and $(\Delta+\hat{P})$ (acting on traceless tensors):
\be
&&\hat{D}_{\mu\nu}^{\ \ \alpha\beta}=(\Delta -
\frac{R}{6})\, \delta_{\mu\nu}^{\alpha\beta}+2W^{(\alpha \ \beta )}_{\ (\mu \ \nu)}\, ,\nonumber \\
&&(\Delta+\hat{P})_{\mu\nu}^{\ \ \alpha\beta}=(\Delta -
C\frac{R}{4})\, \delta_{\mu\nu}^{\alpha\beta}+aW^{(\alpha \ \beta )}_{\ (\mu \ \nu)}\, .
\lb{50}
\ee
We see that  for the choice of parameters
\be
C=\frac{2}{3}\, , \ \ a=2\, 
\lb{51}
\ee
we have that $\hat{D}=(\Delta +\hat{P})$ and the  non-locality in (\ref{49}) disappears for any Einstein metric. The quadratic action then takes a local ghost-free form
\be
\delta_2W=-\frac{M^2_{eff}}{2}\int \frac{1}{4}\bar{h}^{\mu\nu}\hat{D}\bar{h}_{\mu\nu}\, .
\lb{52}
\ee
This is in agreement with the findings in \cite{Barvinsky:2011hd}.
The analysis of ref.\cite{Barvinsky:2011hd} was concentrated mostly on the maximally symmetric de Sitter space (round sphere $S^4$) and the condition on the parameter $C$
(\ref{51}) was derived to guarantee the absence of the ghosts in the quadratic action over this maximally symmetric background. The condition on the parameter $a$ (\ref{51}) is new and it guarantees the absence of the ghosts 
in the quadratic action over any Einstein spacetime. Thus, despite the non-local and higher derivative nature of the action (\ref{22}), it does not describe any ghosts or extra degrees of freedom and propagates only that of the standard spin-2 particle.  

It is interesting to note that the quadratic theory (\ref{52}) can be considered as a ``square root'' of the higher-derivative theory (\ref{7}), discussed  in section 2.

\medskip

\noindent {\it Solution  with zero cosmological constant, $\Lambda=0$.} Let us briefly discuss the Ricci flat solution, $R_{\mu\nu}=0$, that corresponds to the 
zero cosmological constant in (\ref{1}). In this case the auxiliary field $\phi$ satisfies the homogeneous equation $\Delta\phi=0$. Demanding the regularity of the field everywhere including  infinity we obtain that the only solution possible is constant, $\phi=const$. The same conclusion is made for the field $\psi$. The traceless tensors $\tilde{\psi}_{\alpha\beta}$ and $\tilde{\phi}_{\alpha\beta}$ satisfy the homogeneous equation $(\Delta +\hat{P})\tilde{\psi}_{\alpha\beta}=0$, where on a Ricci flat background the matrix $\hat{P}$ contains only the Weyl term, $\hat{P}_{\alpha\beta}^{\ \ \mu\nu}=aW_{(\alpha\ \beta)}^{\ (\mu \ \nu )}$. In flat spacetime the arguments using equation (\ref{38})
show that, demanding the regularity at infinity, the only solution of this equation is such that $\nabla_\alpha\tilde{\psi}_{\mu\nu}=0$, i.e.  $\tilde{\psi}_{\mu\nu}$ should be constant. In a curved spacetime a priori there is no covariantly constant traceless tensor so that a possible solution is trivial, $\tilde{\psi}_{\mu\nu}=0$. Whether this solution is unique in the presence of a  non-vanishing Weyl tensor is not evident.  We will  not dwell into this problem in the present paper. In any case,  there exists the following 
solution\footnote{It should be noted that the original non-local model of gravity described by the action (\ref{22}) is equivalent to a local theory with the auxiliary fields provided the certain boundary conditions are imposed on the auxiliary fields (otherwise the new formulation may have more degrees of freedom than the original one). In the asymptotically flat spacetime the Green's function of operator $(\Delta-\alpha R)$ is decaying at infinity. Therefore the function $\phi=(\Delta-\alpha R)^{-1}R$ vanishes at the asymptotic infinity. In the local formulation, however, any constant value of $\phi$ is possible. We remark that only the zero value $\phi=0$ is compatible with the original
non-local formulation of the theory.}   
\be
R_{\mu\nu}=0\, , \ \  \ \ \tilde{\psi}_{\mu\nu}=0\, , \ \ \phi=const\, , \ \ \psi=0\, , \ \ \tilde{\phi}_{\mu\nu}=0\, ,
\lb{53}
\ee
where we choose the trivial solution for $\psi$, that describes a Ricci flat spacetime.

The Ricci flat spacetime may contain some horizons (such as the horizon in the Schwarzschild metric). The horizon entropy  
is still given by the equation (\ref{entropy}). For the solution
(\ref{53}) one then finds  that the entropy 
\be
S=4\pi\int_\Sigma(1+\alpha\phi)^2\, 
\lb{54}
\ee
is proportional to the area of the horizon as in  General Relativity.  This entropy is non-vanishing provided the constant $\phi$ is such that $(1+\alpha\phi)$ is non-zero.

For a small perturbation $h_{\mu\nu}$ over a  background (\ref{53}) we find for the quadratic part of the action
\be
\delta_2 W=\int\left(-(1+\alpha\phi)(\delta_2R+\frac{1}{2}\Tr h \, \delta_1 R)-\alpha \phi\delta_1 R\delta_1 \phi+4\alpha \delta_1E^{\mu\nu}\delta_1\tilde{\psi}_{\mu\nu}\right)\, .
\lb{55}
\ee
In the gauge $\nabla_\alpha h^\alpha_\beta=\frac{1}{2}\partial_\beta \Tr h$ we have that
\be
&&\delta_2 R=\frac{1}{4}\bar{h}^{\mu\nu}\hat{D}\bar{h}_{\mu\nu}+\frac{3}{16}\Tr h\, \Delta \Tr h\, , \nonumber \\
&&\delta_1R=-\frac{1}{2}\Delta \Tr h\, , \ \ \delta_1 \phi=(1+\alpha\phi)\frac{1}{\Delta}\delta_1 R\, , \nonumber \\
&&\delta_1 E_{\mu\nu}=-\frac{1}{2}\hat{D}\bar{h}_{\mu\nu}\, , \ \ \delta_1 \tilde{\psi}_{\mu\nu}=\frac{1}{\Delta+\hat{P}}\delta_1 E_{\mu\nu}\, ,
\lb{variations}
\ee
where $\bar{h}_{\mu\nu}=h_{\mu\nu}-\frac{1}{4}g_{\mu\nu}\Tr h$ is the traceless part of $h_{\mu\nu}$ and $\delta_1\phi$  and $\delta_1\tilde{\psi}_{\alpha\beta}$ are obtained by varying equations (\ref{29}) and (\ref{33}) respectively.
Operator  $\hat{D}$ in (\ref{variations}) is the same operator as in (\ref{47}) (with $R_{\mu\nu}=0$). We remind that $\phi=const$ in (\ref{55}) and (\ref{variations}).
Substituting relations (\ref{variations}) into (\ref{55}) we find for the quadratic action
\be
\delta_2W=-(1+\alpha\phi)\int \left(\frac{1}{4}\bar{h}^{\mu\nu}\hat{D}\bar{h}_{\mu\nu}-\frac{(1-4\alpha)}{16}\Tr h\, \Delta \Tr h\right)   +\alpha\int \bar{h}^{\mu\nu}   \hat{D}\frac{1}{(\Delta+\hat{P})}\hat{D}\bar{h}_{\mu\nu}\, .
\lb{56}
\ee
For a  Ricci flat metric the operators $\hat{D}$ and $(\Delta +\hat{P})$ (\ref{50}) coincide if parameter $a=2$ (no condition on constant $C$ is required in this case).
For this value of $a$ the non-locality in (\ref{56}) disappears and the quadratic 
action (\ref{56}) then takes the local form
\be
\delta_2W=-\frac{M^2_{PL}}{2}\int \left(\frac{1}{4}\bar{h}^{\mu\nu}\hat{D}\bar{h}_{\mu\nu}-\beta(\alpha,\phi)\frac{1}{16}\Tr h \, \Delta \, \Tr h\right)\, .
\lb{57}
\ee
The Planck mass in this phase is
\be
\frac{M^2_{Pl}}{2}=(1+\alpha\phi-4\alpha)\, ,
\lb{58}
\ee
and we introduced parameter $\beta(\alpha,\phi)$ as follows
\be
\beta(\alpha,\phi)=\frac{(1+\alpha\phi)(1-4\alpha)}{(1+\alpha\phi-4\alpha)}\, .
\lb{beta}
\ee
For any value of $\phi$ and small $\alpha$ we have that $\beta(\alpha,\phi)=1+O(\alpha^2)$. On the other hand, for $\phi=0$ we have that $\beta(\alpha,\phi=0)=1$ for any value of $\alpha$. We note that only the zero value $\phi=0$ is compatible with the original non-local formulation of the theory (\ref{22}) (see the footnote 6.)
 For $\beta(\alpha,\phi)=1$ the quadratic action (\ref{57}) is exactly the quadratic part of the Einstein-Hilbert action (with zero cosmological constant)
\be
\delta_2 W=-\frac{M^2_{PL}}{2}\int \delta_2 (R\sqrt{g})\, .
\lb{action}
\ee
For $\phi=0$ this agrees  with \cite{Barvinsky:2011hd}, \cite{Barvinsky:2003kg}. Our analysis shows that there are no ghosts in  the phase with $\Lambda=0$. For small $\alpha$ the gravitational coupling in this phase is much stronger than the coupling (\ref{48})  in the phase with a non-zero cosmological constant. 

\medskip

\noindent {\it Two phases of the theory.} We see that in the model (\ref{22}) there exist two rather different phases: with non-vanishing cosmological constant $\Lambda>0$ and with
$\Lambda=0$. The horizon entropy in these two phases, as we have shown,  is radically different: zero in the phase with $\Lambda>0$ and proportional to the area in a standard way in the phase with $\Lambda=0$. Moreover, these two phases are characterized by two different gravitational couplings (see equations (\ref{48}) and (\ref{58})). 
The source of the difference between these two phases lies in the  equation (\ref{29-1}) for the auxiliary field $\phi$. In the Euclidean spacetime with $\Lambda>0$ the only solution
to this equation is $\phi=-1/\alpha$. The quadratic term in the variation of  the Ricci scalar  in action (\ref{local})  is multiplied by $(1+\alpha\phi)$ and, thus, does not contribute to the quadratic action. On the other hand, in the phase $\Lambda=0$ the only solution of (\ref{29-1})  for $\phi$, compatible with the original non-local formulation of the theory (\ref{22}),   is $\phi=0$. In this case we find that the quadratic term in $R$ does contribute to the quadratic action as in General Relativity.

It should be noted that these two phases are not analytically related and are realized on two different spacetimes\footnote{We understand that our interpretation of the phases is different from the one presented in \cite{Barvinsky:2011hd}.} (with $R>0$ and $R=0$ respectively). The discontinuity between these two phases is eventually related to the different topological properties of asymptotically flat space-time and
the de Sitter spacetime, in the Euclidean signature the latter is a closed manifold while the former has a boundary at infinity  where the extra boundary conditions should be imposed.
 The presence of a non-vanishing cosmological constant is visible in  all scales.
No matter how small or large is black hole the horizon entropy is zero in this case. At all scales, the coupling of matter to a weak gravitational field is governed by the coupling  constant (\ref{48}). This is contrary to, perhaps rather naive, expectations that on a smaller scales than cosmological (say, near the Earth) one could effectively use, in a weak field approximation,    the quadratic action (\ref{57}) with the gravitational coupling (\ref{58}). (An example of  the model when the gravitational coupling does depend on the scale is considered in \cite{Parikh:2000fn}.) Of course, it is possible that in the model (\ref{22}) there could exist a domain wall solution which connects the two phases (the Ricci scalar then would not be   constant) with zero and non-zero values of $\Lambda$. It would be interesting to see whether the model (\ref{22}) indeed admits a solution of this type. 

\section{Conclusion}

In the class of theories considered in this paper the problem of statistical explanation of the entropy associated to horizons becomes trivial.
Since the entropy is zero the corresponding quantum  state is  a single pure state. In a description in terms of a 2d conformal field theory 
of \cite{CFT} this  may correspond to a situation when the corresponding central charge vanishes. The CFT then does not contain any non-trivial degrees of freedom. We note that this trivialization of the problem is somehow related to the presence of a positive cosmological constant.
Indeed, in the non-local model of Barvinsky there exists a phase with zero cosmological constant, where the horizon entropy is proportional to the area as
in the standard situation of General Relativity. Only the presence of any, even extremely tiny, positive cosmological constant changes the situation radically
so that the entropy becomes zero. We note that this is so for any size of horizon: microscopical or cosmological. Clearly, this applies also to the largest horizon possible, to the  cosmological horizon in de Sitter spacetime.  This property  is, apparently,  due to the non-local nature of the model.

One might worry that the property of vanishing of the horizon entropy in the discussed class of theories is due to  some kind of pathology hidden  in the theory.
At first sight this hypothetical pathology may be related to the non-unitarity and the appearance of the ghosts.  However, the presence of ghosts does not seem to be a necessary and unavoidable feature of the discussed theories. The non-local model studied in section 3.2 is free of ghosts, at least in the quadratic order, for any value of $\Lambda$. 

Based on the observations made in this paper it is tempted to speculate that the presence of a positive cosmological constant should play a crucial role in the resolution of the so-called information puzzle. This however should be taken with a great care since any horizon present in the spacetime (\ref{1}), semi-classically, still emits the  Hawking radiation. How this can be  
compatible with the described property of the entropy remains to be understood.

\bigskip

\section*{Acknowledgements} 

This project has started when the author was visiting the group of high energy and cosmology  at the  University of Munich. The kind hospitality 
of G. Dvali is greatly acknowledged. The author thanks A. Barvinsky for the stimulating discussions 
and E. Humbert for a useful discussion of the
Lichnerowicz operator. I thank A. Barvinsky for reading the manuscript and the useful remarks.

\setcounter{equation}0

\end{document}